\documentclass[12pt]{iopart}

\expandafter\let\csname equation*\endcsname\relax
\expandafter\let\csname endequation*\endcsname\relax
\usepackage{amstext}
\usepackage{amssymb}
\usepackage{graphicx}
\usepackage{dcolumn}
\usepackage{bm}
\usepackage{subfigure}
\usepackage{xcolor}
\usepackage{cite}
\usepackage{amsfonts}
\usepackage{amssymb}
\usepackage{amsmath}
\usepackage{txfonts}
\usepackage{bm}
\usepackage{dsfont}
\usepackage{tikz}
\usepackage{graphicx}

\usepackage{isomath}

\begin{document}

\title{Quantifying  Dynamical Total Coherence in a Resource Non-increasing Framework}
\author{Si-ren Yang$^1$, Chang-shui Yu*$^{1,2}$}

\begin{abstract}
We quantify the dynamical  quantum resource  in the resource non-increasing (RNI) framework, namely,  the free dynamical  resource is defined by the channels that cannot increase the  static ``resourcefulness'' of any input state.  We  present  two kinds of approaches to quantifying the dynamical resource, the distance measures and  the maximal increasing static resource (MISR).  As a demonstration, we quantify the dynamical  total coherence with our presented measures. It is shown that the distance based measures have good operational interpretation through quantum processing tasks and can be numerically calculated by semidefinite programming (SDP) and the measures based on MISR could lead to the analytical solution.  As an application,  we consider the dynamical total coherence of the qubit amplitude damping channel. Both the analytical measure based on the static $l_2$ norm and the numerical illustrations based on the SDP are given. 
\end{abstract}

\address{$^1$School of Physics, Dalian University of Technology, Dalian 116024,
China}
\address{$^2$DUT-BSU joint institute, Dalian University of Technology, Dalian 116024, China }
\ead{ycs@dlut.edu.cn} \vspace{10pt}
\begin{indented}
\item[\today]
\end{indented}

\vspace{2pc} \noindent\textit{Keywords}: best convex approximation, quantum state preparation, quantum
coherence
%
%
%
\maketitle
\section{Introduction}

In recent years, there has been increasing interest in  quantum technology which extensively develops our recognization of quantum world.
 Quantum coherence originating from quantum superposition   displays the departure between quantum world and classical world and has  been shown to be closely related to quantum  entanglement \cite{entanglement, entangle,e1,e5}, quantum discord \cite{discord,qd1,qd2,qd3,qd4}, quantum asymmetric \cite{as,ass1,piani,as3} and other features. Quantum coherence also plays a crucial role in many quantum processes such as quantum algorithms \cite{grover,qa1,qa2}, quantum computation \cite{qco1,qco2,qco3,qco4}, and quantum thermodynamics \cite{ther1, ther2, ther3, ther4, ther5, ther6}.  Since T. Baumgratz et. al. \cite{Quantifying}  proposed the framework of quantum resource theory (QRT) to quantify coherence under specific constraints and pre-requests, many researches have studied the  coherence of quantum states especially on coherence measures \cite{alter,trace,multi,mea1,mea2,mre,mea3} and operational interpretation \cite{op1,op2,op3,op4,op6}. 
 
 The coherence of quantum states belong to the static QRT which has two ingredients: the free states without any resource of interest, and the free operations which cannot convert the free states to those with resource. For example,  in the QRT of coherence, the free states are the incoherent states which are diagonal in  some fixed basis, and the free operations are the incoherent operations which will never transfer an incoherent state into a coherent state. In this sense, many other quantum features such as quantum entanglement \cite{entanglement}, nonlocality \cite{nonlocality}, contextuality \cite{contextuality}, non-Gaussianity \cite{Strobel424}, asymmetry \cite{as} and so on  \cite{sp1, sp2, usp1,usp2} have also been widely investigated. In particular, the basis-free coherence such as total coherence  \cite{2016Total} has been proposed and the operational meaning of the total coherence has also been found   \cite{YANG2018305} . 
  
Besides the static resources, researchers concern more and more on dynamical systems, i.e., quantum channels, which have an obvious motivation that most quantum information processes are dynamical and accomplished by quantum channels.  Dynamical systems carry  higher dimensional information, which implies that quantum features in quantum channels can also be taken regard as resources. Therefore, the dynamical QRT \cite{ch3,ch5,POVM1,qrt5,ch9,ch8,ch1,qrt1,nch1,nch2,nch3,nch4,qd,nch5,nch6}could also been established similar to but quite different from the QRT in static systems. Ref. \cite{ch3} quantifies the coherence of a channel with the corresponding Choi state, and study coherifications of quantum channels. 
In Ref. \cite{ch5}, a dynamical  resource theory based on completely positive-Wigner-preserving (CPWP) quantum operations as free operations, quantum CPWP superchannels as superoperators is introduced and  two efficiently computable magic measures are provided. In Ref. \cite{POVM1}, the resource theory is built in terms of the free Positive-Operator-Valued Measures (POVMs) and free detection/creation incoherence in the sense of quantum computational setting \cite{qrt5}. In Ref. \cite{ch9}, the dynamical coherence is investigated for Gaussian channels. In addition, the dynamical entanglement is also considered in Ref. \cite{ch8}, and a quantitative connection between the coherence and entanglement on the level of operations are introduced in Ref. \cite{ch1}. All the relevant researches are based on the resource non-generating (RNG)  \cite{qrt1} or  its upgraded  completely resource non-generating (CRNG) frameworks, namely, the free operations are defined as those that cannot generate any resource from free resource (even assisted by an ancillary reference).

In this paper, we propose the dynamical resource theory in a different perspective, namely, define the dynamical QRT in the resource non-increasing (RNI) framework. If the RNI is only constrained on the free states subject to a static resource, the RNI framework will be exactly the RNG framework. 
To establish the QRT, we also consider  the superoperations in  the sense of quantum computing pre-requests analogous to similar to Ref. \cite{ch1}. The alternative candidates have been presented to quantify dynamical  resources under the RNI framework and  the quantification of the total quantum coherence has been deeply studied as one demonstration. We construct two types of measures of dynamical total coherence and find the operational meaning of the measure of dynamical total coherence by the quantum channel discrimination tasks. As examples, the analytical and the semidefinite programming (SDP) \cite{sdp} are provided and the calculations of concrete measures of the dynamical total coherence are also  demonstrated with regard to the  qubit amplitute damping channel.

The remaining of this paper is organized as follows. In section II,  we review the ingredients of the dynamical QRT and present the new dynamical QRT in the RNI framework.  In Section III,  we present a few measures for the dynamical total coherence  based on the distance and the maximal increasing static resource, then provide the analytical expression and the SDP for the proposed measures and  give their concrete calculations for the qubit amplitude damping channel. We summarize all the conclusions in Section V.

\section{Resource theory of quantum channels}
Quantum dynamic resource is referred to as the process of dynamical evolution and  is equally described by quantum channels which  are the completely positive trace preserving (CPTP) maps. In this paper, we denote such quantum channels with capital Greek letters($\mathrm{\Theta}$, $\mathrm{\Phi}$, $\mathrm{\Psi}$, etc) mapping a density operator in Hilbert space $\mathcal{H}^A$ to another state in $\mathcal{H}^B$.
To investigate the ``resourcefulness'' of quantum channels, we have to establish QRT for them. Similar to the  static QRT of quantum states which typically includes two ingredients: free operations and free states \cite{op1,qrt1},  here we also need two basic ingredients of QRT of quantum channels: free dynamical resources $\mathcal{F}$ and free superoperations $\mathfrak{F}$.  However,  different from the free resource of the well-studied  RNG framework, we introduce our  free dynamical ingredient by the following definition.

 \textit{Definition 1-} Let $M(\rho)$ denote a well-defined static resource measure for the given state $\rho$ (in the sense of  Ref. \cite{Quantifying}), the channel $\mathcal{F}$ is free if 
\begin{equation}
\mathsf{\Delta} M(\mathcal{F})=\displaystyle  \max_{\rho \in \mathcal{D}}\, \{M(\mathcal{F}(\rho))-M(\rho)\}\leq 0
\end{equation}
with $\mathcal{D}$ representing the set of all density matrices. 

For example, if we consider the 
 dynamical QRT of quantum coherence, one alternative definition of the free dynamical coherence can be given by substituting $M(\rho)$ with proper static coherence measures $C(\rho)$ \cite{Quantifying}.
It is obvious that the free operations defined above will not exceed the maximally free operations which are only constraint with free static resource.
It is not difficult to understand in that  the ``resourcefulness'' is relative to some quantum processes with certain operational meaning.
This naturally leads to that  the ``resourcefulness'' based on different static resource measure are incompatible with each other, which is similar to the widely-known incompatibility of  ordering two states subject to different measures of the same resource. Of course, to avoid such an compatibility, one can always conceptually maximize $\mathsf{\Delta} M(\mathcal{F})$ over all the potential $M$, which will, nevertheless, lose the computability generally. In addition, the above definition  indicates that our free channels would not increase any ``resourcefulness'' of a given state, which is why we called  it  RNI.  It is obvious that  the free operations in the RNI framework are also free in the RNG framework.

 Another ingredient in QRT of dynamical resource is the set of free superchannels which is a subset of the space of supermaps.  The space of supermaps is a vector space equipped with inner products that would map one linear map to another. Thus, superchannels would map a channel to another, and they are  described by linear maps which are  CP preserving and TP preserving. There are several ways to realize a superchannel such as  with pre- and post- processings \cite{gstein}, and  with several free elements \cite{qo} . Here we make a further extension of Ref. \cite{qo}  as follows. 

 \textit{Definition 2.-} \cite{qo} A superoperation given in the Kraus representation $\{ \mathfrak{F}_n\}$ is free if and only if $\mathfrak{F}_n[\mathrm{\Theta}]$ can be written into the sequence as
 \begin{equation} \mathfrak{F}_n[\mathrm{\Theta}]=\mathcal{E}_{i_n,\mathrm{\Phi}^{j_n}_n},\cdots,\mathcal{E}_{i_2,\Phi^{j_2}_2}\mathcal{E}_{i_1,\mathrm{\Phi}^{j_1}_1},\end{equation} 
 where $\mathcal{E}_{i_n,\mathrm{\Phi}^{j_n}_n}$ denotes the $j_n$th Kraus element of the superoperation $\{\mathcal{E}_{i_n,\mathrm{\Phi}^{j_n}_n}\}$ with the corresponding free operation $\{\mathrm{\Phi}^{j_n}_n\}$ and 
 \begin{align}
 & \mathcal{E}_{i_n=0,\mathrm{\Phi}^{j_n}_n}[\mathrm{\Theta}]=\mathrm{Tr}[\mathrm{\Theta}], \\
 &\mathcal{E}_{i_n=1,\mathrm{\Phi}^{j_n}_n}[\mathrm{\Theta}]=\mathrm{\Phi}^{j_n}_n\circ \mathrm{\Theta},\mathcal{E}_{i_n=2,\mathrm{\Phi}^{j_n}_n}[\mathrm{\Theta}]=\mathrm{\Phi}^{j_n}_n\otimes \mathrm{\Theta},  \\
 &\mathcal{E}_{i_n=3,\mathrm{\Phi}^{j_n}_n}[\mathrm{\Theta}]=\mathrm{\Theta\circ\Phi}^{j_n}_n,\mathcal{E}_{i_n=4,\mathrm{\Phi}^{j_n}_n}[\mathrm{\Theta}]=\mathrm{\Theta\otimes \Phi}^{j_n}_n.
 \end{align}

From Definition 2, one can find that the superoperation with $i_n=0$ will erase all the information of the input state and finally assign the maximally mixed state (identity). The superoperations with $i_n=1,3$ correspond to the additional free operation $\mathrm{\Phi}^{j_n}_n$ after or before the operation $\mathrm{\Theta}$, while the superoperations with $i_n=2,4$ mean that the free operation $\mathrm{\Phi}^{j_n}_n$ is performed on an auxiliary system attached on the system of interest. With these superoperations,  a dynamical QRT can be immediately established so long as we specify a free dynamical resource. It is obvious that free dynamical resources are the channels that do not possess the feature of interest, and free superoperations are the superchannels  that can only map free channels to free channels. In this sense,  our free  superoperations can be given in Kraus representation as $\mathfrak{F}=\sum_n p_n \mathfrak{F}_n$, if $\mathfrak{F}_n[\mathcal{F}]\in S $ for any $\mathcal{F}\in S$, where $S$ denotes the corresponding set of free operations.

With the two  fundamental RNI ingredients of QRTs, we would like to quantify quantum features of the dynamical resources. 
By means of some functions applied on the quantum channel, dynamical measures obtains a scalar which represents  the ``resourcefulness'' of the channel. In particular, this applied function should satisfy the axiomatic requirements of a QRT, which can be given in the following rigorous way \cite{qrt1}. 

\textit{Definition 3.-} Any functional  measure $T(\cdot)$ quantifying  dynamical resource of  arbitrary quantum channel ( denoted by $\mathrm{\Theta}$)  is a qualified measure if 
 \begin{align}\nonumber
(1)&\text{ Faithfulness}: T\mathrm{ (\Theta)} \geq 0, \text{with } '=' \text{ holds  iff } \mathrm{\Theta}\in S \text{ is free};\\ \nonumber
(2)&\text{ Monotonicity or Strong monotonicity}: T(\mathrm{\Theta}) \geq T(\mathfrak{F}[\mathrm{\Theta}]) \\ \nonumber 
&\text{or} \ T(\mathrm{\Theta})\geq\sum_{n}p_{n}T(\mathrm{\Theta}_{n}) \text{ for free superoperations } \\ \nonumber & \mathfrak{F}=\sum_{n}p_{n}\mathfrak{F}_{n} \text{ with  } 
 \mathrm{\Theta}_{n}=\mathfrak{F}_{n}\mathrm{[\Theta]}\: \text{and}  \sum_n p_n=1,\\\nonumber
(3)&\text{ Convexity}:  T(\mathrm{\Theta}) \text{ is convex}.\nonumber
 \end{align}

The strong monotonicity is an operational constraint where the measures obey monotonicity under selective measurements. However, if one could not have enough information about a superoperation such as a black box in practical scenarios,  it is enough to consider the monotonicity, even though in a QRT the strong monotonicity is required. In addition,  the free superoperations can also be explicitly given in the sense of quantum computational setting similar to Ref.  \cite{qo}, namely, a quantum operation can be plugged in the network.

The methods to quantifying a dynamical resource are not unique. Here in order to establish the dynamical QRT of coherence, one can quantify a given operation by the deviation from  Definition 1 or quantify the distance of a given operation from the set of free operations. Our suggested measures could be given in the following form.

 \textit{Definition 4.-} Dynamical coherence of any quantum channel $\mathrm{\Theta}$ can be measured as
\begin{align}
T(\mathrm{\Theta})&=\displaystyle \min_{\mathcal{F}\in  S_C} \vert\vert \mathrm{\Theta} -\mathcal{F}\vert\vert_{\star}, \label{normmeasure}\\
\tilde{T}(\mathrm{\Theta})&=\displaystyle \max \{\mathrm{\Delta} C_{\star}(\mathrm{\Theta}), 0\},\label{minus}
\end{align}
where $S_C$ denotes the set of free operation with regard to coherence, and the subscript $\star$ emphasizes any well-defined distance norm in Eq. (\ref{normmeasure}) or static coherence measure in Eq. (\ref{minus}) (if it exists) in the sense of  Definition 3.

We would like to emphasize that Definitions 3 and 4 mainly take  coherence as examples to elucidate our dynamical QRT in terms of the coherence non-increasing (CNI) framework. \textit{If the coherence measure $C(\cdot)$ in both definitions is replaced by other well-defined quantum resources, the dynamical QRT corresponding to the particular resources can be established in a corresponding quantum resource non-increasing (RNI) framework. Moreover, the RNI framework can be generalized to the  completely RNI (CRNI) similar to CRNG versus RNG by admitting tensor product structure \cite{qrt1}}.  

\section{ Dynamical Total Coherence}

Quantum total coherence is one of quantum resources in quantum theories. It is a basis independent  coherence and can be regarded as one kind of static resource.  It has been illustrated its operational meaning in Ref. \cite{YANG2018305}. Operational resource theory of  total coherence has a direct fact that the only free (incoherent) state is the maximally mixed state $\frac{\mathds{1}_A}{\vert A\vert} $ with $A$ denoting the dimension since  all potential bases or observables have been considered. The free operations  are selective unitary operations that would not change the coherence of the free state.  Based on Definition 3, in order to establish the dynamical QRT of the total coherence, we will employ a good measure of the static total coherence measure.  It is worthy mentioned that  quantum total coherence has a close relation with purity. Notably, in Ref. \cite{GOUR20151}, authors investigated ``nonuniformity" which is also related to static purity. Here we focus on the dynamical case.

As shown in Ref. \cite{2016Total},  the  relative entropy  $C_{RE}(\rho)=\mathrm{log}\ n-S(\rho)$ is one of good measures for the static total coherence of a given density matrix  $\rho$.  
Thus the free operations with respect to the relative entropy can be given in the following theorem.

\textit{Theorem 5.-} Unital Channels $\{\mathcal{F_U}(\mathds{1}) =\mathds{1},\mathcal{F_U}\in \text{CPTP}\}$ are the free operations  under CNI framework (Definition 3) with the relative entropy as static measures for total  coherence.

\textit{Proof. }(Sufficiency.) The quantum channel $ \mathcal{E}$ can be expressed in Kraus representation as $\mathcal{E}(\cdot)= \sum_{i}{E}_{i}(\cdot){E}_{i}^{\dagger}$ with its Kraus operators $\sum_{i} {E}_{i}^{\dagger}{E}_{i}=\mathds{1} $.  Therefore, unital channels $ \mathcal{F_U}$ have its Kraus operators $ \textit{F}_{i}$ satisfy $ \sum_{i} \textit{F}_{i}^{\dagger} \textit{F}_{i}=\sum_{i} \textit{F}_{i}\textit{F}_{i}^{\dagger}=\mathds{1}$. For arbitary input state $\rho \in \mathcal{D}$ (the linear space of density matrix) going through a unital channel, its output state $\sigma=\mathcal{F_U}(\rho)$ will be majorized by the input state, i.e. $\rho\succ\sigma$ \cite{major}. Obviously,  the output's total coherence measured by relative entropy will never exceed the input state. Thus the set of unital channels belongs to total coherence non-increasing channel. 

(Necessity.)  We notice that  total coherence non-increasing channels $ \mathcal{R}$ should not increase coherence when we input the only  incoherent state $ \frac{\mathds{1}_{A}}{\vert A\vert}$. This means the output state would be the maximally mixed state itself. Eliminate  dimension of the system, we have $\mathcal{R}(\mathds{1})=\mathds{1}$, which is the definition of unital channels.  Combine the sufficiency and necessity, we can conclude that unital channels are total coherence non-increasing channels.   $\Box$

\textit{Remark.- } (a) Unital channels have different subclasses, the mixed unitary channels we utilized  in the static resource theory of total coherence is just one of them. But in the situation of qubit system, all unital channels are mixed unitary channels \cite{unital}.
(b) Identity channel will map identity to identity which means identity channel is unital and free.

\subsection{Measures based on distance}

From the perspective of distance measure, we choose two widely applied norms: the induced trace norm \cite{unital}defined as 
\begin{align}
\vert\vert\mathrm{\Phi}\vert\vert_{1}=\text{max}\{{\vert\vert\mathrm{\Phi(X)}\vert\vert_{1}:\;X\in\mathcal{L}(X),\: \vert\vert X\vert\vert_{1}\leq 1}\},
\end{align} 
and the completely bounded trace norm \cite{unital} defined as
\begin{align}
\vert\vert\mathrm{\Phi}^{\mathrm{A}\rightarrow\mathrm{B}}\vert\vert_{\diamond}=\vert\vert\mathrm{\Phi}^{\mathrm{A}\rightarrow\mathrm{B}}\otimes\mathds{1}^{\mathrm{C}}\vert\vert_{1},
\end{align}
where $
\vert\vert X\vert\vert_{1}=\text{Tr}(\sqrt{X^{\dagger}X})
$
 for linear operator $X$ in its linear space $\mathcal{L}(X)$,  and  system A and C have the same dimension. In this sense, one can quantify the dynamical total coherence as follows.

 \textit{Theorem 6-} Dynamical total coherence of any quantum channel $\mathrm{\Phi}$ can be measured by  \begin{align}
T _{N}(\mathrm{\Phi})=\displaystyle \min_{\mathcal{F}\in S_{TC}} \vert\vert \mathrm{\Phi} -\mathcal{F}\vert\vert_{N},
 \end{align}
where  $S_{TC}$ is the set of free operations in the sense of the total coherence. In principle, $N$
can denote any norm on quantum operations with the properties $\left\Vert \mathrm{\Theta}_1\circ \mathrm{\Theta}_2\right\Vert\leq\left\Vert  \mathrm{\Theta}_1\right\Vert\left\Vert\mathrm{\Theta_2}\right\Vert$ and $\left\Vert  \mathrm{\Theta}_1\otimes \mathrm{\Theta}_2\right\Vert\leq\left\Vert  \mathrm{\Theta}_1\right\Vert\left\Vert\mathrm{\Theta_2}\right\Vert$ for any given two operations $ \mathrm{\Theta}_1$ and $ \mathrm{\Theta}_2$. Here we especially let $N=1$ and $N=\diamond$ denote the induced trace norm and the completely bounded trance norm, respectively. 

Firstly, one can easily find that the faithfulness is naturally satisfied due to the distance norm and the optimization on the free operations. 

Secondly, it can be shown that the present measure also satisfies the convexity.  Consider the probabilistic mixture of $m$ quantum channels $\mathrm{\Theta}_{m}$ with probability 
$p_{m}$ as $ \mathrm{\Theta}=\sum_m p_m  \mathrm{\Theta}_m$, one can write the average dynamical coherence as $\bar{T}_N=\sum_mp_{m}T_{N}({\mathrm{\Theta}_{m}})=\sum_mp_m \min_{\mathcal{F}_m\in S_{TC}}\left\Vert \mathrm{\Theta}_m-\mathcal{F}_m\right\Vert_N\geq  \min_{\mathcal{F}_m\in S_{TC}}\left\Vert\sum_mp_m \mathrm{\Theta}_m-\sum_mp_m\mathcal{F}_m\right\Vert_N\geq  \min_{\mathcal{F}\in S_{TC}}\left\Vert \mathrm{\Theta}-\mathcal{F}\right\Vert_N=T_N\left( \mathrm{\Theta}\right)$, which is the convexity.

Thirdly, the strong monotonicity also holds. Note that the free superoperation can be written in the Kraus representation as $\mathfrak{F}=\sum_{n}p_{n}\mathfrak{F}_{n}$ with every component given by the free operations as $\mathfrak{F}_n[ \mathrm{\Theta}]=\mathcal{E}_{i_n,\mathrm{\Phi}^{j_n}_n},\cdots,\mathcal{E}_{i_2,\mathrm{\Phi}^{j_2}_2}\mathcal{E}_{i_1, \mathrm{\Phi}^{j_1}_1}$. Since the norms $\left\Vert\cdot\right\Vert_N$ employed here are sub-multiplicative and sub-multiplicative with respect to tensor products, one can immediately have $T_N\left(\mathcal{F}_n( \mathrm{\Theta})\right)\leq T_N\left( \mathrm{\Theta}\right)$, which is analogous to those in Ref. \cite{ch1}. Thus
\begin{align}
\sum_{n}p_{n}T_N(\mathfrak{F}_{n}[ \mathrm{\Theta}])&\leq\sum_{n}p_{n}T_N( \mathrm{\Theta})=T_N( \mathrm{\Theta}).\end{align}

Finally, the strong monotonicity and the convexity give the monotonicity, which completes the proof. $\Box$

Since the above measures  originate from the trace norm, they naturally inherit the operational meaning of the trace norm. Namely, they can be understood by the discrimination tasks in the sense of some \textit{optimal} state undergoing two considered channels, which is demonstrated in the Appendix. 
In addition, we also present SDP \cite{sdp} method to evaluate the measure $T_{\diamond}$ in the Appendix.
In fact, the above norms are not the only choice for the measure of dynamical coherence, the relative entropy is also a good choice, which has been deeply studied in Ref. \cite{qd} for dynamical resource theory. 

\textit{Theorem 7.-} Dynamical total coherence of any quantum channel $\mathrm{\Phi}$ can be measured by:
 \begin{align}
T_{RE}(\mathrm{\Phi})=\displaystyle\inf_{\mathcal{F}\in S_{TC}} \: D(\mathrm{\Phi}\vert\vert\mathcal{F})
\end{align}
with the channel divergence for two quantum channels $\mathcal{N,M}$ defined by
\begin{align}
D(\mathcal{N}\vert\vert\mathcal{M}):=\displaystyle \max_{\phi_{RA}\in \mathcal{D}(RA)}\:S[\mathcal{N}_{A\rightarrow B}(\phi_{RA})\vert\vert\mathcal{M}_{A\rightarrow B}(\phi_{RA})],
\end{align}
where $S(\rho\vert\vert\sigma)=\mathrm{Tr}[\rho\mathrm{log}\rho-\rho\mathrm{log}\sigma]$ is the relative entropy and $\phi_{RA}$ is the pure state in $\mathcal{D}(RA)$. 

\textit{Proof.}  It is trivial to see that   $T_{RE}(\mathrm{\Phi})=0$ if and only if $\mathrm{\Phi}\in \mathcal{F}$. In addition,  the joint convexity of quantum relative entropy directly gives the convexity of  $T_{RE}$.   As to the strong monotonicity, one can consider a free superoperation $\mathfrak{F}=\sum_{n}p_{n}\mathfrak{F}_{n}$ as the proof of Theorem 6, one can easily show
$T_{RE}(\mathfrak{F}_n(\mathrm{\Phi}))\leq T_{RE}(\mathrm{\Phi}) $ due to the monotonicity of relative entropy \cite{qd}. So $\sum_{n}p_{n}T_{RE}(\mathfrak{F}_n(\mathrm{\Phi}))\leq T_{RE}(\mathrm{\Phi}) $, which is exactly the monotonicity. 
The monotonicity can be naturally obtained by the strong monotonicity and the convexity. $\Box$

It also has a close relation with  quantum  Stein's Lemma \cite{stein1,stein2} for the task of discriminating between $n$ copies of a fixed channel $\mathrm{\Phi}$ and one free channel $\mathcal{F}$, which gives a bound on the error probability and hence endows the operational meaning to $T_{RE}$ (see the Appendix).

\subsection{Measure based on the maximal increasing static coherence}
Now we  introduce the measures for dynamical total coherence  in the sense of the increasing static coherence described by Eq. (5). To do so, we will have to select a well-defined static total coherence measure.  However, not all valid static measures can contribute  good computability even in the qubit case. Here we'd like to employ the total coherence measure based on $l_2$ norm, which  is shown in Ref.  \cite{YANG2018305} to be a good quantifier of the total quantum coherence and different from the basis-dependent case. Considering the definition of the $l_2$ norm measure 
\begin{align}
C_{2}(\rho)=\mathrm{Tr}\ \rho^2-\frac{1}{n},
\end{align}
one can explicitly present the coherence of any quantum channel $\mathrm{\Theta}$  based on Eq. (5)  as follows.

\textit{Theorem 8.-} Dynamical total coherence of any quantum channel $\mathrm{\Theta}$ can be measured by 
\begin{align}
\tilde{T}_{2}(\mathrm{\Theta})=\max \{\mathrm{\Delta} C_{2}(\mathrm{\Theta}), 0\},
\end{align}
where $\mathrm{\Delta} C_{2}(\mathrm{\Theta})= \max_{\rho \in \mathcal{D}} \mathrm{Tr}[\mathrm{\Theta}(\rho)^2-\rho^2]$.

\textit{Proof.}   It  is obvious that $\tilde{T}_{2}(\mathrm{\Theta})=0$ for $(\mathrm{\Theta})\in \mathcal{F}$. For convexity, one can consider the superchannel $ \mathrm{\Theta}=\sum_m p_m  \mathrm{\Theta}_m$ as Theorem 6.  Due to the quadratic function of $l_2$ norm, one can find $\tilde{T}_{2}(\sum_m p_m  \mathrm{\Theta}_m) \leq \max_{\rho \in \mathcal{D}}\sum_m p_m\mathrm{Tr} \{\mathrm{\Theta}_m(\rho)^2- \rho^2\}\leq \sum_m p_m\max_{\rho_m \in \mathcal{D}}\mathrm{Tr}\{\mathrm{\Theta}_m(\rho_m)^2-\rho_m^2\}=\sum_m p_m  \tilde{T}_{2}(\mathrm{\Theta}_m)$. 

For the strong monotonicity, one will have to consider a free superoperation $\mathfrak{F}=\sum_{n}p_{n}\mathfrak{F}_{n}$. Since the free operations for the total coherence is the mixed unitary channels, it is easy to find that  $l_2$ norm is also sub-multiplicative and sub-multiplicative with respect to tensor products for any free element $\mathrm{\Phi}_{i} [\mathrm{\cdot} ]$. Thus we have
$\tilde{T}_{2}(\mathfrak{F}_n(\mathrm{\Phi}))\leq \tilde{T}_{2}(\mathrm{\Phi}) $, which implies  $\sum_{n}p_{n}\tilde{T}_{2}(\mathfrak{F}_n(\mathrm{\Phi}))\leq \tilde{T}_{2}(\mathrm{\Phi}) $. 
The monotonicity is automatically given by the strong monotonicity and the convexity. $\Box$

This definition has the same motivation as measures based on relative entropy under RNI or CRNI framework but gives a different value. One obvious advantage of the measure in Theorem 8 is that the measure can be analytically calculated in some cases, which is demonstrated in the following.

\textit{Theorem 9.-} Given a 2-dimensional quantum channel  $\mathrm{\Theta}(\cdot)=\sum_{\alpha} K_{\alpha}(\cdot)K_{\alpha}^{\dagger}$ with  $a_{i}=\frac{1}{2}\mathrm{Tr}\mathrm{\Theta}(\sigma_{i})$
,  $M_{ij}=\frac{1}{2}\mathrm{Tr}\sigma_{i}\mathrm{\Theta}(\sigma_{j})$, the dynamical total coherence reads 
\begin{align}
\tilde{T}_{2}(\mathrm{\Theta})=\mathrm{max}\{\sum\limits_{i=1}^3\frac{\xi_{i}^2\tilde{a}^2_{i}}{2(1-\xi^{2}_{i}))}+\frac{a^2_{i}}{2},\ 0\},
\end{align}
where $\xi_{i}$ is the singular value of the matrix $M$,  $\vert\tilde{a}\rangle={U^{T}}\vert a \rangle$ with $U$ determined by the singular value decomposition ${M=U\Lambda V^{\mathrm{T}}}$.

\textit{Proof. } The proof is given in the Appendix. $\Box$

\subsection{Examples}
As applications, we consider the dynamical total coherence of  amplitude damping channel of a single qubit \cite{nielsen}.  The amplitude damping channel  is  a fundamental quantum channel describing the energy dissipation in quantum process with a damping parameter $\eta$. The Kraus operators for this channel read
\begin{align}
&K_{0}=
\left[
\begin{array} {lr}
1 & 0\\
0&\sqrt{1-\eta}\\
\end{array}
\right],
&K_{1}=
\left[
\begin{array} {lr}
0& \eta\\
0&0\\
\end{array}
\right].
\end{align}
\begin{figure}
\centering
\includegraphics[scale=0.5]{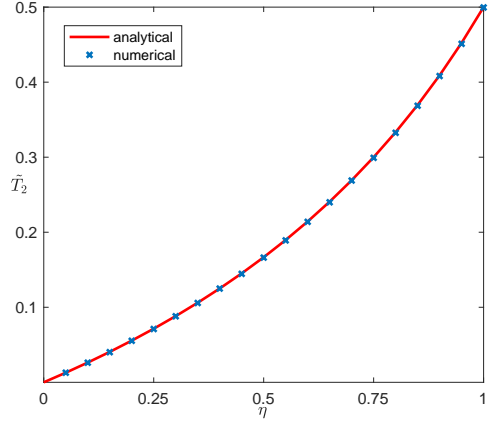}
\caption{The dynamical total coherence  by $\tilde{T}_2$  versus $\eta$.}\label{T12}
\end{figure}
Fig. \ref{T12} shows the dynamical total coherence of amplitude damping channel measured by $\tilde{T}_2$ versus $\eta$. When the damping parameter is 0, the amplitude damping channel is free, and its total coherence is increasing with the  parameter $\eta$. This is consistent with the fact that when the parameter $\eta$ equals 0, the amplitude channel becomes the identity channel that is free. While $\eta$ reaches one, it is easy to verify this channel would map all input states to the ground state which is a resourceful state in the total coherence framework, and obviously, the channel is  resourceful at this time.  In addition, In Fig. \ref{T12} we have numerically test the results by optimizing among randomly generated density matrices, which are completely consistent with the analytical results.

\begin{figure}
\centering
\includegraphics[scale=0.5]{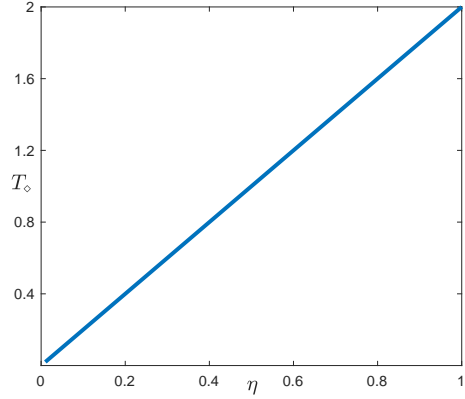}
\caption{The dynamical total coherence $T_{\diamond}$  versus $\eta$.}\label{TD} 
\end{figure}

\begin{figure}
\centering
\includegraphics[scale=0.5]{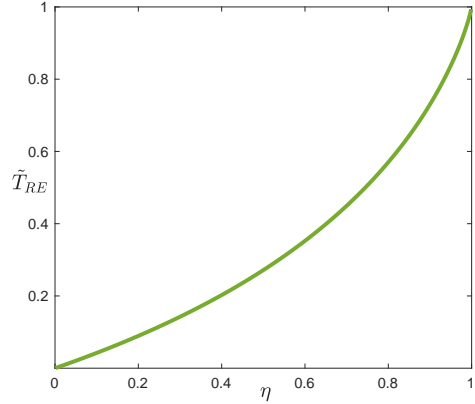}
\caption{The dynamical total coherence $\tilde{T}_{RE}$ versus $\eta$.}\label{TRE}
\end{figure}

The measure $T_{\diamond}$ for the amplitude damping channel is demonstrated in Fig. 2, where one should note that the free operations are based on the relative entropy of total coherence. In addition, as a comparison, we also give try another measure $\tilde{T}_{RE}$ defined as
 \begin{align}
\tilde{T}_{RE}(\mathrm{\Theta})=\displaystyle \max \{\mathrm{\Delta} C_{RE}(\mathrm{\Theta}), 0\}
\end{align}
in  Fig. \ref{TRE}, where the subscript $RE$ means that the relative entropy is selected to quantify the static coherence.
 Compared with  $T_2$, we can find that the dynamical total coherences with different measures have the same tendency with the damping parameter $\eta$, but they show an obviously different understanding on the free operations.  It is worth mentioning that both  $T_{\diamond}$ and $\tilde{T}_{RE}$ are subject to the relative entropy as static total coherence, so they not only have a similar tendency, but more importantly the two lines in Figs. \ref{TD},\ref{TRE} have the same initial point. That is,   $T_{\diamond}$ and $\tilde{T}_{RE}$ represent two different methods to quantifying the dynamical total coherence, but they should belong to the same frameworks (have the same definitions of free operations). This shows  that under the same static measure, both distance measure and maximal variance are  effective to evaluate the ``resourcefulness'' of a quantum channel.

\section{Discussion and Conclusion}
In this paper, we have investigated how to quantify dynamical resources under a new proposed resource theory framework, i.e., RNI and CRNI frameworks. These frameworks are  motivated by that free dynamical resources could not increase the ``resourcefulness'' for arbitrary input static resources.  The RNI framework gives a new insight into the RNG dynamical resource theory  and fulfills the requirements of quantum resource theory. The  distances and  the maximal increasing static resource are suggested to quantify  the dynamical resource with the  precondition that the free operations are defined based on  a well-defined static resource measure. As the main demonstration, we use the relative entropy of total coherence to define the free operations and employ the induced trace norm and the diamond norm  to quantify the dynamical total coherence, which implies  good operational meanings. In addition, we employ the $l_2$ norm based total coherence measure to quantify the dynamical total coherence in the sense of the maximal increasing static resource, which can be analytically calculated for any qubit quantum channel. 
As examples, we calculate the dynamical total coherence of the amplitude damping channel based on the analytical and the  SDP numerical approaches.
\section*{ACKNOWLEDGEMENTS}
The authors thank Gilad Gour, Carlo Maria Scandolo, Yun-long Xiao, Jian-wei Xu for inspiring discussion.  This work was supported by the National Natural Science Foundation of China under Grant
No. 11775040, No. 12011530014, the Fundamental Research Fund for the Central Universities under Grant No. DUT20LAB203, and the Key Research and Development Project of Liaoning Province under Grant No.2020JH2/10500003.

\appendix

\section{Operational meaning of $T_N$}
In the quantum channel discrimination task \cite{discrimination2, qsl1,qsl2, qsl3}, Alice and Bob are two participants. Alice prepares a probability state in a classical register Z, which has a probability $\lambda$ in state 0 and probability $1-\lambda$ in state 1. X and Y are two arbitrary registers. Alice receives the register X from Bob who prepared an initial state $\sigma$ in it. Conditioned on the classical state of Z, Alice performs two actions: (1) if Z=0, transforms X into Y according to channel $\mathcal{N}_1$; (2) if Z=1, transforms X into Y according to channel $\mathcal{N}_2$ which is belong to  free channels. Bob's goal is to determine the classical state Z by only applying subset of free operations. His strategy  is to maximize the quantity $\frac{1}{2}\vert\vert\lambda\rho_{1}-(1-\lambda)\rho_{2}\vert\vert$ with the output states $\rho_{1}$, $\rho_{2}$ coming from $\mathcal{N}_1$, $\mathcal{N}_2$ respectively \cite{unital}.

Here, we  set a scenario with  $\lambda=\frac{1}{2}$ for simplicity, and allow Bob physically to hold an auxiliary register R.  We will show some approaches for Bob who wants  to improve the probability to discriminate two different quantum channels. Now, the probability for Bob succeeding to discriminate two channels is \cite{disprop,qo}
\begin{align}
\text{Prob}(\mathcal{N}_1,\mathcal{N}_{2})=\frac{1}{2}+\frac{1}{4}\displaystyle \max_{\mathrm{\Psi}^{\text{AR}},\sigma^{AR}}\vert\vert\mathrm{\Psi}^{\text{AR}}(\mathcal{N}_{1}-\mathcal{N}_{2})^{A}\otimes\mathds{1}^{\text{R}}\sigma^{\text{AR}}\vert\vert_1,\label{proba}
\end{align}
where $\mathrm{\Psi}^{\text{AR}}$ are allowed additional operations performed by Bob, he would decide perform it or not (i.e. identity operation which is  belong to $S_{TC}$). In particular, if Bob holds only free channels $\mathcal{F}^{AR} \in S_{TC}$ and we minimize the induced diamond distance among $\mathcal{N}_2\in S_{TC}$, one can easily see that
\begin{align}\nonumber
 T_\diamond\left(\mathcal{N}_{1}\right)&=\displaystyle \min_{\mathcal{N}^{A}_2 \in S_{TC}} \vert\vert\mathcal{N}^{A}_1-\mathcal{N}^{A}_2\vert\vert_\diamond\\\label{maxf}
& \geq \displaystyle \min_{\mathcal{N}^{A}_2 \in S_{TC}}\displaystyle \max_{\mathcal{F}^{AR}\in S_{TC}} \vert\vert\mathcal{F}^{AR}(\mathcal{N}_1-\mathcal{N}_2)^{A}\vert\vert_\diamond\\\nonumber
&=\displaystyle \min_{\mathcal{N}^{A}_2 \in S_{TC}}\displaystyle \max_{\mathcal{F}^{AR}\in S_{TC}} \vert\vert\mathcal{F}^{AR}(\mathcal{N}_1-\mathcal{N}_2)^{A}\otimes \mathds{1}^{R}\vert\vert_1\\\label{tra}
&=\displaystyle \min_{\mathcal{N}^{A}_2 \in S_{TC}}\displaystyle \max_{\mathcal{F}^{AR}\in S_{TC},P} \displaystyle \max_{\sigma^{AR}}\text{Tr}[P\mathcal{F}^{AR}(\mathcal{N}_1-\mathcal{N}_2)^{A}\otimes \mathds{1}^{AR}\sigma^{AR}]\\\nonumber
&=\displaystyle \min_{\mathcal{N}^{A}_2 \in S_{TC}}\displaystyle \max_{\mathcal{F}^{AR}\in S_{TC},\sigma^{AR}}\vert\vert\mathcal{F}(\mathcal{N}_1-\mathcal{N}_2)^{A}\otimes \mathds{1}^{AR}\sigma^{AR}\vert\vert_1\\\label{absorb}
\end{align}
where (\ref{maxf}) comes from the monotonicity of dynamical measures. (\ref{tra}) is due to the trace norm $\left\Vert\cdot\right\Vert= \max_P \text{Tr} [P (\cdot)]$ with $P$ denoting the projective operator. Since any projective operators belong to $S_{TC}$, thus we have (\ref{absorb}).

Contrast (\ref{absorb}) with (\ref{proba}), one can easily see that in such a scenario the probability is exactly given by
 \begin{equation}
 \text{Prob}(\mathcal{N}_1,\mathcal{N}_{2})=\frac{1}{2}+\frac{1}{4}\displaystyle T_\diamond\left(\mathcal{N}_{1}\right),\label{hh}
 \end{equation}
 since the maximum can be reached at least once when Bob do an identity operation. Hence, the boundary gives the operational meaning to $T_\diamond$.

If we don't consider the auxiliary reference R, one will directly find the operational meaning of the dynamical measure $T_1$ based on Eq. (\ref{hh}), namely, one can directly replace the diamond norm in Eq. (\ref{hh}) by the usually trace norm.

In addition, one can understand $T_1$ in other settings. If  Bob knows the basis of initial prepared state, he can coincidently recognize the basis of initial  bipartite state $\sigma^{RA}$ which can be written as $\sigma^{RA}=\sum_{i,i'}p_{i,i'}\vert\phi_{i}, i\rangle \langle\phi_{i'},i' \vert$ in reference bases $\vert \phi_{i} \rangle$ and $ {\vert i\rangle}$. Then he can apply a dephasing  channel $\mathsf{\Delta}(\cdot)=\sum_{i}\langle i\vert\cdot\vert i\rangle \vert i\rangle \langle i\vert$ on the auxiliary register R and any unital channel on register A, then send it to Alice. In the end, Bob can discriminate $\mathcal{N}_1$, $\mathcal{N}_2$ with probability no more than the coherence of two channels:
\begin{align}\label{wat}
&\text{Prob}_{dp}(\mathcal{N}_1,\mathcal{N}_{2})\\\nonumber
=&\frac{1}{2}+\frac{1}{4}\displaystyle \max_{\sigma^{AR}}\vert\vert(\mathcal{F}^{\text{A}}\otimes \mathsf{\Delta}^{R})(\mathcal{N}_{1}-\mathcal{N}_{2})^{A}\otimes\mathds{1}^{\text{R}}\sigma^{\text{AR}}\vert\vert_1\\\nonumber
=&\frac{1}{2}+\frac{1}{4}\displaystyle \max_{\sigma^{AR}} \vert\vert(\mathcal{F}^{\text{A}}\otimes \mathsf{\Delta}^{R})(\mathcal{N}_{1}-\mathcal{N}_{2})^{A}\otimes\mathds{1}^{\text{R}}\sum_{i}p_{i,i'}\vert\phi_{i}, i\rangle \langle\phi_{i'},i' \vert\vert_1\\\nonumber
= &\frac{1}{2}+\frac{1}{4}\displaystyle \max_{p_{i,i},\vert\phi_{i}, i\rangle}\vert\vert \sum_{i}p_{i,i}\mathcal{F}^{A}(\mathcal{N}_{1}-\mathcal{N}_{2})^{A}
\vert\phi_{i}\rangle \langle\phi_{i}\vert^{A}\otimes\vert i \rangle \langle i\vert ^{R}\vert\vert_1\\\nonumber
=& \frac{1}{2}+\frac{1}{4}\displaystyle \max_{\sigma^A} \vert\vert\mathcal{F}^{A}(\mathcal{N}_{1}-\mathcal{N}_{2})^{A}
 \sigma^{A}\vert\vert_1\\\nonumber
=& \frac{1}{2}+\frac{1}{4}\vert\vert(\mathcal{N}_{1}-\mathcal{N}_{2})\vert\vert_1,\end{align}
where the subscript 'dp' means that Bob first performs a proper dephasing operation and $\sigma^A=\sum_i p_{i,i}\left\vert\phi_i,i\right\rangle\left\langle\phi_i,i\right\vert$.

Similarly, if we consider all $\mathcal{N}_2 \in {S}_{TC}$, one will immediately find that 
\begin{equation} 
\text{Prob}_{dp}(\mathcal{N}_1,\mathcal{N}_{2})=\frac{1}{2}+\frac{1}{4}\displaystyle \max_{\mathcal{N}_2 \in \mathcal{F}_{TC}} \vert\vert\mathcal{N}_{1}-\mathcal{N}_{2}\vert\vert_1=\frac{1}{2}+\frac{1}{4} T_1(\mathcal{N}_1).\label{hha}
\end{equation}

In more general situations,  Bob  cannot collect  information about the basis of the initial state, we consider it as a basis independent problem and can optimize the probability with basis independent quantum resource. Bob can prepare a free maximally mixed state in the auxiliary register, and plays free interaction with the final state. The probability would be
\begin{align}
&\text{Prob}(\mathcal{N}_1,\mathcal{N}_{2})\\\nonumber
=&\frac{1}{2}+\frac{1}{4}\displaystyle \max_{\sigma^{A}}\vert\vert\mathcal{F}^{\text{AR}}(\mathcal{N}_{1}-\mathcal{N}_{2})^{A}\otimes \mathds{1}^{\text{R}}(\sigma^{\text{A}}\otimes \frac{\mathds{1^{R}}}{\vert R\vert})\vert\vert_1\\\nonumber
=&\frac{1}{2}+\frac{1}{4}\displaystyle \max_{\sigma^{A}}\vert\vert(\mathcal{N}_{1}-\mathcal{N}_{2})^{A}\sigma^{A}\otimes (\frac{\mathds{1^{R}}}{\vert R\vert})\vert\vert_1\\\nonumber
=&\frac{1}{2}+\frac{1}{4} \displaystyle \max_{\sigma^{A}}\vert\vert(\mathcal{N}_{1}-\mathcal{N}_{2})^{A}\sigma^{A}\vert\vert_1\cdot\vert\vert\frac{\mathds{1^{R}}}{\vert R\vert}\vert\vert_1\\\nonumber
=& \frac{1}{2}+\frac{1}{4}\vert\vert(\mathcal{N}_{1}-\mathcal{N}_{2})\vert\vert_1,
\end{align}
which reaches the same result as Eq. (\ref{wat}) and will lead to the same conclusion as Eq. (\ref{hha})
 
 \section{Operational meaning of  $T_{RE}$}

$T_{RE}$ has a close relation with  Quantum  Stein's Lemma \cite{stein1,stein2} for quantum channels. In the task of  identifying  resourceful channel $\mathcal{N}\in\mathcal{H}_{A}\rightarrow \mathcal{H}_{B}$ and free channel $\mathcal{M}_{n}\in\mathcal{F}_{A^{n}\rightarrow {B^{n}}}$, we would like to emphasize the difference from channel discrimination task  that  the observer holds n realizations of the resourceful channel and one free channel. The observer applies a free initial state $\rho_{RA}\in \mathcal{D}(RA)$ and measures a two-outcome positive operator value measure (POVM) $\{A_{n},I-A_{n}\}$. He would claim the channel is $\mathcal{N}$ if obtains the outcome is associated to $A_{n}$ or claim the channel is $\mathcal{M}_{n}$ if the outcome is associated to $I-A_{n}$. However, there exist two types of errors:

$\bullet\:$ Type I: The observer guesses the channel is free while it  is  $\mathcal{N}\in\mathcal{H}_{A}\rightarrow \mathcal{H}_{B}$. This happens with probability $\alpha_{n}(A_n):=\mathrm{Tr}[\mathcal{N}^{\otimes n}_{A\rightarrow B}(\rho^{\otimes n}_{RA})(I-A_{n})]$.

$\bullet$ Type II: The observer guesses the channel is resourceful while  it belongs to  $\mathcal{M}_{n}\in\mathcal{F}_{A^{n}\rightarrow {B^{n}}}$. This happens with probability $\beta_{n}(A_{n}):=\mathrm{Tr}[\mathcal{M}_{n} (\rho^{\otimes n}_{RA})A_{n}]$. 

In the test, the type II error should be minimized to the extreme, while only requiring  that the probability of type I is bounded by a small parameter $\epsilon$. We define the minimal type II error probability under type I parameter as (see related research \cite{qd,qsl1,qsl2,qsl3,gstein}):
\begin{align}
\beta_{n}(\epsilon, A_{n}):=\displaystyle \min_{0\leq A_{n}\leq I} \{\beta_{n}(A_{n}): \alpha_{n}(A_{n})\leq \epsilon\}
\end{align}

According to Quantum Stein's Lemma,  for every $0\leq\epsilon\leq 1$,
 \begin{align}
\displaystyle \lim_{n\rightarrow \infty}-\frac{\mathrm{log}(\beta_{n}(\epsilon, A_{n}))}{n}=S(\sigma_{\mathcal{N}}\vert\vert\sigma_{\mathcal{M}}),
\end{align}
where $\sigma_{\mathcal{N}}$ and $\sigma_{\mathcal{M}}$ are \textit{ single shot} output states of $\mathcal{N}$ and $\mathcal{M}$.

The observer will maximize the divergence between $\mathcal{N}$  and $\mathcal{M}$   which also minimizes the probability of type II overall  input states:
\begin{align}
\displaystyle \max_{\rho\in\mathcal{D}(RA)} \displaystyle \lim_{n\rightarrow \infty}-\frac{\mathrm{log}(\beta_{n}(\epsilon, A_{n}))}{n}=\displaystyle \max_{\rho\in\mathcal{D}(RA)} S(\sigma_{\mathcal{N}}\vert\vert\sigma_{\mathcal{M}})
\end{align}

In general situations, the observer does not have information about the identity of free channel, he only picks one free channel from the free domain. The loss of information makes the optimal programming to minimize the probability overall the free channels: 
\begin{align}
\displaystyle \min_{\mathcal{M}\in\mathcal{F}}\displaystyle \max_{\rho\in\mathcal{D}_{RA}} \displaystyle \lim_{n\rightarrow \infty}-\frac{\mathrm{log}(\beta_{n}(\epsilon, A_{n}))}{n}=\displaystyle \min_{\mathcal{M}\in\mathcal{F}}\displaystyle \max_{\rho\in\mathcal{D}_{RA}} S(\sigma_{\mathcal{N}}\vert\vert\sigma_{\mathcal{M}}).
\end{align}
Thus, we can conclude that the minimal error that occurs in such a quantum dynamical identifying task is bounded by the channel's dynamical total coherence $T_{RE}$.

\section{Calculation of $T_{\diamond}$  by  SDP}
For arbitrary quantum channel $\mathcal{E}$, it can be expressed by Choi representation (sometimes referred to as Choi-Jamiolkowski isomorphism)\cite{CHOI,Jami} as
\begin{align}
J(\mathcal{E})=\mathds{1}\otimes\mathcal{E}(\phi_{+}),
\end{align}
where $\phi_{+}$ is unnormalized maximally entangled state. Its dynamical total coherence can be measured by
\begin{align} 
T_{\diamond}(\mathcal{E})=\displaystyle \min_{\mathcal{F}\in \mathrm{FREE}}\vert\vert \mathcal{E}-\mathcal{F}\vert\vert_{\diamond}
\end{align} 
and evaluated by semidefinite programming (with polynomial algorithms \cite{sdp1}). According to its definition, diamond norm for operator $\vert\vert \mathcal{E}-\mathcal{F}\vert\vert_{\diamond}$ has its primal problem
\begin{align}
&\textit{Primal}\\\nonumber
\mathrm{minimize} \quad &2\vert\vert\mathrm{Tr(Z)}\vert\vert_{\infty}\\\nonumber
\textit{s.t.}\quad &\mathrm{Z}\geq J(\mathcal{E}-\mathcal{F})\\\nonumber
&\mathrm{Z}\geq 0.
\end{align}
Then $T_{\diamond}(\mathcal{E})$ is the optimal value of 
\begin{align}
\mathrm{minimize}:\quad &2\vert\vert\mathrm{Tr(Z)}\vert\vert_{\infty}\\\nonumber
\textit{s.t.}\quad \mathrm{Z}&\geq J(\mathcal{E}-\mathcal{F})\\\nonumber
\mathrm{Z}&\geq 0\\\nonumber
\mathcal{F}&\:\mathrm{is\: unital}.
\end{align}

Applying constraints on $\mathcal{F}$ and Choi representation properties for unital channels, the primal problem is:
\begin{align}
\mathrm{minimize}:\quad &2\vert\vert\mathrm{Tr(Z)}\vert\vert_{\infty}\\\nonumber
\textit{s.t.}\quad \mathrm{Z}&\geq J(\mathcal{E})-W\\\nonumber
&\mathrm{Z}\geq 0\\\nonumber
&\mathrm{W}\geq0\\\nonumber
&\mathrm{Tr_{B}(W)}=\mathds{1}_{A}\\\nonumber
&\mathrm{Tr_{A}(W)}=\mathds{1}_{B}.\nonumber
\end{align}
This is equivalent to 
\begin{align}
\mathrm{minimize:}\: a\\\nonumber
\textit{s.t.} \quad &a\geq0\\\nonumber
&\mathds{1}_{A}\cdot a-2\mathrm{Tr_{B}(Z)}\geq 0\\\nonumber
& \mathrm{Z}\geq J(\mathcal{E})-W\\\nonumber
&\mathrm{Z}\geq 0\\\nonumber
&\mathrm{W}\geq0\\\nonumber
&\mathrm{Tr_{B}(W)}=\mathds{1}_{A}\\\nonumber
&\mathrm{Tr_{A}(W)}=\mathds{1}_{B}.\nonumber
\end{align}
The Lagrangian of primal problem is given by: 
\begin{align}
\mathcal{L}(a,Z,W,\widetilde{X},X,Y_1,Y_2)=a+\mathrm{Tr}[(2\mathrm{Tr_{B}(Z)}-a\mathds{1}_{A})\widetilde{X}]\\\nonumber
+\mathrm{Tr}[(J(\mathcal{E})-W-Z)X]+\mathrm{Tr}[(\mathrm{Tr_{A}W-\mathds{1}_{B}})Y_{1}]\\\nonumber
+\mathrm{Tr}[(\mathrm{Tr_{B}W}-\mathds{1}_{A})Y_{2}]
\end{align}
and its dual function (see more details for prime and dual problems in Ref. \cite{unital}) is 
\begin{align*}
&q(\widetilde{X},X,Y_{1},Y_{2})=\displaystyle \inf_{a,Z,W } \mathcal{L}(a,Z,W,\widetilde{X},X,Y_1,Y_2)\\\nonumber
&\qquad=\displaystyle \inf_{a,Z,W } \mathrm{Tr}[J(\mathcal{E})X]-\mathrm{Tr}[Y_1+Y_2]+a[1-\mathrm{Tr}(\widetilde{X})]\\\nonumber
&+\mathrm{Tr}[(2\cdot\widetilde{X}\otimes\mathds{1}_{B}-X)Z]+\mathrm{Tr}[(\mathds{1}_{A}\otimes Y_{1}+Y_{2}\otimes \mathds{1}_{B}-X)W].\nonumber
\end{align*}
The dual function has value $\mathrm{Tr}[J(\mathcal{E})X]-\mathrm{Tr}[Y_1+Y_2]$ if $ \widetilde{X}\leq 1\land 2\cdot\widetilde{X}\otimes\mathds{1}_{B}-X\geq0\:\land\mathds{1}_{A}\otimes Y_{1}+Y_{2}\otimes\mathds{1}_{B}-X\geq0$ and  $-\infty$ in other cases. Thus, the dual problem is to maximize $\mathrm{Tr}[J(\mathcal{E})X]-\mathrm{Tr}[Y_1+Y_2]$  with constraints for $\widetilde{X}, X, Y_1,Y_2$. To make the constraint simple and easy reading, we conclude the four constraints: (1-2) $\widetilde{X}$ is nonnegative  and trace less than one; (3) $X \geq 0$;  (4) $2\cdot\widetilde{X}\otimes\mathds{1}_{B}-X$ as one, i.e. $X\leq 2\cdot\rho\otimes\mathds{1}_{B}$ which $ \rho$  is a density matrix . Then we can construct a 
$\widetilde{X}^{'}:=\frac{1}{\mathrm{Tr}\widetilde{X}}\widetilde{X}$ is trace one,  positive semidefinite and will keep $2\cdot\widetilde{X}^{'}\otimes\mathds{1}_{B}-X$ positive semidefinite for all X satisfy $2\widetilde{X}\otimes\mathds{1}_{B}-X\geq 0$.  The dual problem can be simplified to:
\begin{align}
&Dual \\\nonumber
\mathrm{maximize}\quad &\mathrm{Tr}[J(\mathcal{E})X]-\mathrm{Tr}[Y_{1}+Y_{2}]\\\nonumber
s.t.\quad& X\leq 2\cdot\rho\otimes\mathds{1}_{B}: \rho\mathrm{\:is\:density\: matrix}\\\nonumber
&\mathds{1}_{A}\otimes Y_{1}+Y_{2}\otimes\mathds{1}_{B}-X\geq0\\\nonumber
&X\geq0\\\nonumber
&Y_{1}=Y_{1}^{\dagger}\\\nonumber
&Y_{2}=Y_{2}^{\dagger}.\nonumber
\end{align}

Finally, we have to show our primal and dual problem reaches the same optimal value, in other words,  the strong duality holds in this programming.
The strong duality obtains if the Slater condition \cite{slater} holds, i.e.there exists $Z^{*}$ and $W^{*}$ satisfy all the equality constraints in primal problem and \textit{strictly} satisfy all the inequality constraints. Obviously, the Slater condition holds with 
\begin{align}
Z^{*}&=\mathds{1}_{A}\otimes\mathds{1}_{B}+J(\mathcal{E})\\\nonumber
W^{*}&=\frac{1}{\vert A\vert\times \vert B\vert}\mathds{1}_{A}\otimes\mathds{1}_{B}.  
\end{align}

\textit{Example.} The quantum channel $\mathcal{K}$ has its Kraus operators:
\begin{align}
&K_{0}=
\left[
\begin{array} {lr}
-0.5084&   -0.5495\\
0.5318&   -0.5108\\
\end{array}
\right],\\
&K_{1}=
\left[
\begin{array} {lr}
0.6701&    0.0846\\
0.0981&   -0.6558\\
\end{array}
\right].
\end{align}
 By solving SDP in software cvx \cite{cvx}, this quantum channel has total coherence as +0.0528434.

\section{Proof of theorem 9}
Let's consider the qubit  dynamical system and show an analytical solution for this specific case. In Bloch representation, any input qubit can be expressed as: $\rho=\frac{\mathds{1}+ \textbf{r}\cdot\bm{\sigma}}{2}$ with  pauli matrices $ \bm{\sigma}=(\sigma_{x}\ \sigma_{y}\ \sigma_{z})$ and its corresponding vector $\textbf{r}=\vert r \rangle=(r_1\  r_2 \ r_3)^{\mathrm{T}}$.  Quantum channel $\mathrm{\Theta}$ has its Kraus representation: $\mathrm{\Theta}(\cdot)=\sum_{\alpha} K_{\alpha}(\cdot)K_{\alpha}^{\dagger}$ and measured by 
$\tilde{T}_2(\mathrm{\Theta})=\displaystyle \max_{\rho\in\mathcal{D}_{2}} \ \mathrm{Tr}\ \mathrm{\Theta}(\rho)^2-\mathrm{Tr}\ (\rho)^2=\displaystyle \max_{\textbf{r}}\ \frac{1}{2}\sum\limits_{i=1}^3(r'^{2}_{i}-r^2_{i}),$
where $\textbf{r}'=(r'_1 \ r'_2 \ r'_3)^{\mathrm{T}}$ is the vector of output qubit state, has its expression as: 
\begin{align}
r'_{i}=a_{i}+\sum\limits_{j=1}^3M_{ij}r_{j}
\end{align}
with
\begin{align}
a_{i}=\frac{1}{2}\sum\limits_{\alpha}\mathrm{Tr}(\sigma_{i}K_{\alpha}K_{\alpha}^{\dagger})
\end{align}
and
\begin{align}
M_{ij}=\frac{1}{2}\sum\limits_{\alpha}\mathrm{Tr}(\sigma_{i}K_{\alpha}\sigma_{j}K_{\alpha}^{\dagger}).
\end{align}
We define a vector $\vert a\rangle=(a_{1}\ a_2\ a_3)^{\mathrm{T}}$ and a matrix $M=\sum_{ij}M_{ij}$ and rewrite $\tilde{T}_2(\mathrm{\Theta})$ as
\begin{align}\nonumber
\tilde{T}_2(\mathrm{\Theta})&=\frac{1}{2}\displaystyle \max_{\textbf{r}} \langle a\vert a \rangle+2\langle a\vert M\vert r\rangle+\langle r\vert (M^{\mathrm{T}}M-\mathds{1})\vert r\rangle\\\nonumber
&=\frac{1}{2}\displaystyle \max_{\textbf{r}} \langle a\vert a \rangle+2\langle a\vert \mathrm{U \Xi V^{T}}\vert r\rangle+\langle r\vert \mathrm{V(\Xi\Xi-\mathds{1})V^{\mathrm{T}}}\vert r\rangle\\\nonumber
&=\frac{1}{2}\displaystyle \max_{\vert \tilde{r}\rangle} \langle a\vert a \rangle+2\langle \tilde{a}\vert \mathrm{\Xi} \vert\tilde{r}\rangle+\langle \tilde{r}\vert(\mathrm{\Xi\Xi}-\mathds{1})\vert\tilde{r}\rangle\\\nonumber
&=\frac{1}{2}\displaystyle \max_{\vert \tilde{r}\rangle} \sum\limits_{i} a^2_{i}+2\sum\limits_{i}\xi_{i}\tilde{a}_{i}\tilde{r}_{i}+\sum\limits_{i}(\xi^2_{i}-1)\tilde{r}^2_{i}\label{T2}.
\end{align}
We apply singular value decomposition on matrix $\mathrm{M=U\Xi V^{\mathrm{T}}}$ in the first equation where U and $\mathrm{V^T}$ are all unitary matrices (eigenvectors of $\mathrm{MM^{T}}$ and $\mathrm{M^{T}M}$ respectively), $\mathrm{\Xi}$ is a diagonal matrix with singular values $\xi_{i}$. In the second equation, we set new vectors $\vert\tilde{a}\rangle=\mathrm{U^{T}}\vert a \rangle$ and $\vert\tilde{r}\rangle=\mathrm{V^T}\vert r\rangle$ are merely related with quantum channel $\mathrm{\Theta}$. However, unitary operators on $\textbf{r}$ would not change the constraint as it should obey in Bloch representation, i.e. : $\sum\limits_{i}\tilde{r}^2_{i}\leq 1$.

 Ignoring the coefficient in the front, we just need to maximize $f(\tilde{r}_{1},\tilde{r}_2,\tilde{r}_3)= \sum\limits_{i} a^2_{i}+2\sum\limits_{i}\xi_{i}\tilde{a}_{i}\tilde{r}_{i}+\sum\limits_{i}(\xi^2_{i}-1)\tilde{r}^2_{i}$ by the method of \textit{Lagrange Multipliers}:
 \begin{align}
\mathcal{L}(\tilde{r}_1,\tilde{r}_2,\tilde{r}_3,\lambda)=f(\tilde{r}_{1},\tilde{r}_2,\tilde{r}_3)+\lambda(\tilde{r}^2_1+\tilde{r}^2_2+\tilde{r}^2_3-1).
 \end{align}

The maximum can be reached when all the partial derivatives equal zero and inequality constraint converts into Karush-Kuhn-Tucker conditions \cite{KKT}:
\begin{equation}\label{lagra}
\left\{
\begin{array} {lr}
\displaystyle \frac{\partial{\mathcal{L}(\tilde{r}_1,\tilde{r}_2,\tilde{r}_3,\lambda)}}{\partial\tilde{r}_{1}}=0,\\
\displaystyle  \frac{\partial{\mathcal{L}(\tilde{r}_1,\tilde{r}_2,\tilde{r}_3,\lambda)}}{\partial\tilde{r}_{2}}=0,\\
\displaystyle \frac{\partial{\mathcal{L}(\tilde{r}_1,\tilde{r}_2,\tilde{r}_3,\lambda)}}{\partial\tilde{r}_{3}}=0,\\
\lambda(\tilde{r}^2_1+\tilde{r}^2_2+\tilde{r}^2_3-1)=0.
\end{array}
\right.
\end{equation}
By solving Eq. (\ref{lagra}), we  get $\tilde{r}$
\begin{equation}
\left\{
\begin{array}{lr}
\tilde{r}_{1}=\displaystyle \frac{\xi_{1}\tilde{a}_1}{(1-\xi^{2}_{1}))} \\
\tilde{r}_{2}=\displaystyle \frac{\xi_{2}\tilde{a}_2}{(1-\xi^{2}_{2}))} \\
\tilde{r}_{3}=\displaystyle \frac{\xi_{3}\tilde{a}_3}{(1-\xi^{2}_{3})} \\
\end{array}
\right.
\end{equation}
to maximize the expected value as dynamical total coherence for every qubit quantum channel:
\begin{align}
\tilde{T}_{2}(\mathrm{\Theta})=\mathrm{max}\{\sum\limits_{i=1}^3\frac{\xi_{i}^2\tilde{a}^2_{i}}{2(1-\xi^{2}_{i}))}+\frac{a^2_{i}}{2},\ 0\}.
\end{align}
in which all  parameters can be calculated conveniently. We conclude that dynamical total coherence measure $\tilde{T}_{2}(\mathrm{\Theta})$ has an analytical solution for qubit channels.

\bibliographystyle{iopart-num}%
\bibliography{ref}

\end{document}